\documentclass[twocolumn,showpacs,preprintnumbers,amsfonts,amsmath,amssymb]{revtex4}
\usepackage{graphics}
\usepackage{epsfig}
\def\ut.#1{\mathop{#1}\limits_{\raisebox{0.4ex}[0pt]{$\sim$}}}
\begin{document}
\title{Bose--Einstein condensate of kicked rotators}
\author{B. Mieck}
\email{mieck@theo-phys.uni-essen.de}
\author{R. Graham}
\email{graham@uni-essen.de}
\affiliation{Department of Physics in Essen,
             University Duisburg--Essen, Universit{\"a}tsstrasse 5,
													45117 Essen, Germany}

\begin{abstract}
A concrete proposal for the realization of a Bose--Einstein condensate of kicked rotators is presented.  Studying their dynamics via the one-dimensional Gross-Pitaevskii equation on a ring we point out the existence of a Lax-pair and an infinite countable set of conserved quantities.
Under equal conditions we make numerical comparisons of the dynamics and their effective irreversibility in time, of ensembles of chaotic classical-, and BECs of interaction-free quantum-, and interacting quantum kicked rotators. \end{abstract}
\maketitle

%

The mechanical system known as `kicked rotator' is an important and well studied minimal model describing the onset of chaos in Hamiltonian dynamical systems. Named after it's original proponents it is also known as `Chirikov-Taylor' map and reads \cite{chir}
\begin{eqnarray}
p_{n+1} &=& p_n + K \sin q_{n+1}\nonumber\\ 
q_{n+1} &=& (q_n + p_n) \mod (2\pi)
\label{eq:1.1}
\end{eqnarray}
The map is generated by the Hamiltonian
$H_R = p^2/2 + K \cos q \sum_n\delta(t-n)$
by taking stroboscopic cross-sections of the dynamics immediately after each kick. $p, q$ is the pair of action angle variables of the system. We use units in which the moment of inertia $\Theta$ and the kicking period $T$ are set to 1. Then the only parameter of the system is $K$. For $K>0.976\dots$ all KAM-curves in phase-space are known to be destroyed \cite{gre} and the system is chaotic. This shows up in a diffusive increase of $\langle p^2_n\rangle$ with $n$ for typical initial ensembles in the $(q,p)$-plane localized in $p$.

This system has also been studied extensively in quantum mechanics, both theoretically \cite{cas,fis,schl} and experimentally \cite{exp}. The quantum system is found to mimick the classical diffusive increase of $\langle p^2\rangle$ only for a finite time interval \cite{cas,fis}. Afterwards the apparent discreteness of the local quasi-energy spectrum of the quantum kicked rotator asserts itself in keeping the expectation value $\langle p^2\rangle$ bounded. A discrete local quasi-energy spectrum implies quasi-energy eigenstates, which are localized in angular-momentum, i.e. square summable over the quantized angular-momentum  $\hbar l$ \cite{fis}. This localization of quasi-energy eigenfunctions implies the boundedness of $\langle p^2_n\rangle$ for all $n$ since an initially localized state has significant overlap only with quasi-energy states localized in its neighborhood. The upshot is that the resulting dynamics of the quantum system is no longer chaotic and instead classifies as quasi-periodic.

An experimental investigation of the quantum kicked rotator was achieved via its quantum optical realization \cite{exp}. There, mixed-state ensembles of laser-cooled noninteracting atoms placed in a periodically pulsed optical lattice could be used to follow in detail the classical and quantum dynamics of noninteracting kicked rotators. However, strictly speaking the quantum optical realization is only made for kicked particles whose configuration space $\mathbb{R}$ is the covering space of the configuration space $\mathbb{R/Z=S}_1$ of the kicked rotator. Classically the dynamics of the kicked particle can simply be projected on the configuration space $\mathbb{R/Z}$ and is then indistinguishable from the dynamics of the kicked rotator. Quantum mechanically the quasi-momentum $\hbar k$ of the non-interacting kicked particle with $0\leq k < 1$ is conserved and equi-distributed in the initial ensemble. It can be eliminated by a linear shift in the momentum $p$, however only at the cost of modifying the periodic boundary condition of the wave function to 
$\psi_k (q+2\pi)= e^{i 2\pi k} \psi_k (q)$.

As a consequence one has realized an ensemble of \underline{different} kicked rotators with boundary conditions labelled by the continuously distributed quasi-momentum $\hbar k$. The total momentum $(p+\hbar k)$ of the kicked particle is not discrete in this ensemble, while the quantized angular momentum $p$ of the kicked rotators is not separately accessible to observation. Averaging over the quasi-momentum is therefore unavoidable and neccessarily
modifies any observed quantum fluctuations.

The advent of Bose--Einstein condensates offers new possiblities:
 
-The new technology to trap and tightly confine ultracold atoms in optical traps now permits to think about new schemes for a direct realization of the quantum kicked rotator, without the detour via the quantum kicked particle.

-
By using an initial state generated by a Bose--Einstein condensate it is now possible to prepare an ensemble of atoms in a pure state. If the condensate is sufficiently diluted in the initial state (e.g. by coherent expansion prior to inclusion in the ring-trap) the subsequent dynamics is still that of interaction-free atoms. However, distinct from the previously realized uncondensed mixed-state ensemble of ultracold atoms, the dynamics of the interacting-free BEC will now have to display the full coherent fluctuations of the quasi-periodic dynamics of the quantum kicked rotator.

-
Taking advantage of the natural interaction in Bose--Einstein condensates the new case of an interacting Bose--Einstein condensate of quantum kicked rotators can also be realized. 

Even prior to the modern interest in Bose--Einstein condensates this system has been studied as an interesting model system \cite{benvenuto}. Recently it has also been studied from the point of view of Bose--Einstein condensation \cite{raizen2}. In particular numerical evidence was given that for sufficiently strong interaction the quasi-periodic dynamics of the quantum-kicked rotator becomes chaotic.

Here it is our aim to make a proposal for the physical realization of the kicked rotator and a corresponding Bose-Einstein condensate with optically trapped ultracold atoms, to derive an infinite number of explicitely time-dependent conserved quantities of the associated Gross-Pitaevskii equation, and to compare the dynamics of the three incarnations of the kicked rotator (classical, quantum without interaction, and quantum with interaction) under otherwise equal conditions.
\begin{figure}
\includegraphics[width=7.3cm, height=4.0cm]{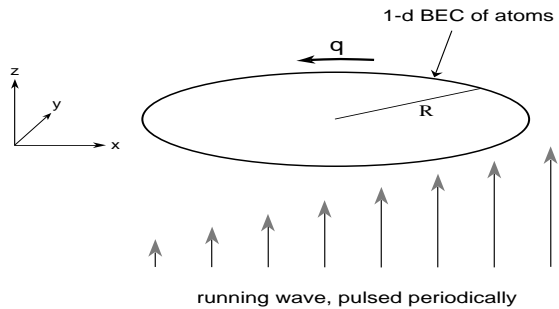}
\caption{\label{fig:1} Proposal for a BEC of kicked rotators. A BEC in a 1-d 
         ring-trap of radius $R$ in the ($x,y$)-plane is illuminated by a 
									periodically pulsed strongly detuned running wave laser in 
									$z$-direction with intensity $I$ engineered to $I=I_0+I_1 
									\frac{x}{R}$. Length of arrows proportional to~$I$.}
\end{figure}

For a physical realization (see fig.\ref{fig:1}) we imagine a Bose--Einstein condensate enclosed in a quasi one-dimensional ring-trap (in the ($x,y$)-plane, say). This can be achieved e.g. by optical trapping of the condensate in an optical  doughnut mode. The angle $q$ of the rotator will be the coordinate $s$ along the ring, $q=s/R$, where $R$ is the radius.  The kicks can be realized \cite{foot}
by the periodically pulsed excitation with a strongly detuned running wave laser transverse to the ring (in the $z$-direction). For the $\cos q$-dependence of the kicking potential, which is proportional to the laser-intensity in this situation, the intensity of the laser beam should be designed such that it increases linearly in one direction (say the $x$-direction with $x=R\cdot\cos q$) in the plane of the ring-trap while being constant across the trap in the two other directions. Thus the periodically pulsed Rabi-frequency $\Omega$ depends only on $x$, $\Omega^2=\Omega^2_0 + \Omega^2_1 \cos q$, and on time $t$. In a way typical for BECs the mean energy must be inferred from the expansion of the cloud of atoms after the trap has been switched off.

Now we turn to the dynamics of the kicked rotator for the interaction-free classical case, the interaction-free quantum case and the interacting quantum case. The first two cases have, of course, been extensively studied in the literature, and are only given here for the sake of comparison.  

The equations numerically solved in the quantum case are
\begin{equation}
ik^{\hspace{-2.3mm}-}\dot{\psi} = -\frac{k^{\hspace{-2.4mm}- 2}} {2}\nabla^2_q\psi + K\cos (q)
\sum^{+\infty}_{n=-\infty}\delta(t-n)\psi + g | \psi |^2 \psi,
\label{eq:1.4}
\end{equation}
with $g=0$, and $g=5$, respectively.
Here time is measured in kicking periods $T$, $k^{\hspace{-2.4mm}-}=\hbar T/\Theta$ with $\Theta = mR^2$ 
is a dimensionless parameter whose size is a measure of the importance of quantum effects. The dimensionless interaction strength is given by 
$g=4\pi NRk^{\hspace{-2.4mm}-}a/Q$ where $Q$ is the effective cross-section of the trap and $a$ is the $s$-wave scattering length. The kicking strength is given by 
$K=(1/\Delta)\int_{\rm{pulse}} \Omega^2_1 (t)dt$, the time-integral over the short pulse in one period, where $\Delta$ is the detuning. It should be noted that the spatially constant $\Omega^2_0$-part of $\Omega^2$ has no physical effect, since it just adds a constant overall phase-shift periodically in time.  (In our 
numerical plots we have chosen the three independent dimensionless parameters $K=5$ well inside the domain of classical chaos, $k^{\hspace{-2.4mm}-} = 0$ or $1$, and  $g = 0$ or $5$.
In all cases we choose as an initial state the one with $p=0$ and $q$ equidistributed over the periodicity interval.)

Before we look at specific numerical solutions we wish to point out a rather remarkable general mathematical structure of (\ref{eq:1.4}) with a surprising additional link to the noninteracting quantum kicked rotator.
This structure reveals itself by the observation that eq.(\ref{eq:1.4}) 
is the one-dimensional non-linear Schr\"odinger equation in an external time-dependent potential (see e.g. appendix D of \cite{abl}). Setting either the external potential or the interaction to zero this equation is integrable. Remarkably, even for nonvanishing interaction and external potential the equation still admitts a Lax-pair, i.e. it can be represented as the compatibility condition
\begin{equation}
\frac{\partial^2 \upsilon}{\partial q \partial t} 
= \frac{\partial^2 \upsilon}{\partial t \partial q}
\label{eq:?.?}
\end{equation}
of the pair of equations
$k^{\hspace{-2.4mm}-}\frac{\partial\upsilon}{\partial q} = \ut.\chi\upsilon$,
$k^{\hspace{-2.4mm}-}\frac{\partial\upsilon}{\partial t} =
\ut.T\upsilon$
with 2x2 matrices
\begin{eqnarray}
  \ut.\chi (q,t) 
&&=        { -ik(q,t)          \qquad  \sqrt{g}\psi^*(q,t)       
 \choose   \!\!\sqrt{g} \psi(q,t) \quad  ik(q,t)},\nonumber\\
\ut.T (q,t)  	
&&=        {A(q,t)            \quad B(q,t)  
 \choose C(q,t)             \quad -A(q,t)}\nonumber
\end{eqnarray}
Here $k(q,t)$ can be chosen in the form 
\begin{equation}
k(q,t)=-ik^{\hspace{-2.4mm}-}\partial\ln \varphi(q,t)/\partial q
\end{equation}
where $\varphi(q,t)$ satisfies the evolution equation
of the {\it noninteracting} quantum kicked rotator, eq.(\ref{eq:1.4}) with
$g=0$. The matrix-elements $A, B,$ and $C$ of $\ut.T$ satisfy three coupled linear inhomogeneous first order partial differential equations in $q$, with coefficients, depending on the external time-dependent potential in (\ref{eq:1.4}) and $\psi, \psi^*, \varphi$ and their first derivatives,
which are all $2\pi$-periodic in $q$. In these equations for $A, B, C$ the time $t$ appears only as a parameter. The detailed explicit form of these equations will not be  needed here beyond the fact that they admit solutions $A, B$ and $C$ which are $2\pi$-periodic in $q$. 

From the Lax-pair, whose existence is often a sign of integrability, a countably infinite set of conserved quantities can be derived: The compatibility conditions (\ref{eq:?.?})  for $\ut.\chi$ and $\ut.T$ implies the relation 
%
 $k^{\hspace{-2.4mm}-} 
( \partial_t \ut.\chi - \partial_q \ut.T ) 
+ [ \ut.\chi, \ut.T ] = 0$
%
%
Defining by a path-ordered integral the 2x2 matrix
\begin{equation}
\ut.U(t) = P \exp \left[
      \frac{1}{k^{\hspace{-2.4mm}-}} \int^{2\pi}_{0} dq
						\ut.\chi (q,t)  \right]
\end{equation}
which is a functional of the condensate wave-function $\psi(q,t)$, $\psi^*(q,t)$, and the external explicitely time-dependent kicking-potential
 at fixed time $t$, and using the $2\pi$-periodicity of $\ut.T(q,t)$ in $q$, $\ut.T(2\pi, t) = \ut.T(0,t)$, it is possible to show that
\begin{equation}
k^{\hspace{-2.4mm}-} \frac{\partial}{\partial t} \ut.U(t)
= \left[\ut.T(0,t)\, ,\,\ut.U(t) \right]
\end{equation}
This implies the existence of infinitely many (explicitly time-dependent) global constants of the motion $C_n ( \{\psi (q,t), \psi^*(q,t) \}, t)=\rm{const}$
with
\begin{equation}
C_n = Tr \ut.U\hspace{-0.1pt}^{n}(t)\, .
\end{equation}
Unfortunately, we have so far not been able to determine the Poisson-brackets of these conservation laws. Conclusions about the integrability of eq.(\ref{eq:1.4}) can therefore not been drawn from this analysis at this point.
At any rate, the dynamics described by (\ref{eq:1.4}) is highly constrained by all these conservation laws. The question whether the dynamics nevertheless
is indeed chaotic for strong coupling, as numerical results of \cite{raizen2}
 suggest, therefore remains a nontrivial one.  It is hard to settle numerically since, as we will see in the following, the numerical error for the strongly interacting system is difficult to
controll over any sufficiently long time-interval. Of course this fact by itself also lends support to the conjecture that the strongly interacting system may be chaotic, the existence of the conservation laws notwithstanding.
\begin{figure}
\includegraphics[width=6.5cm,height=4.0cm]{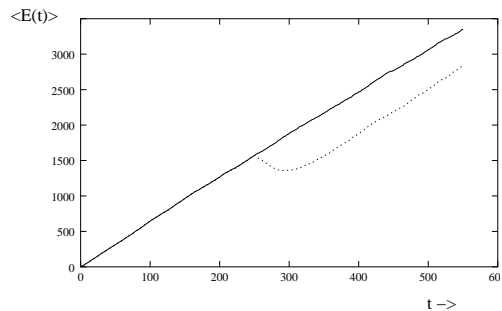}
\caption{\label{fig2} Mean energy $\langle E\rangle$ of the standard map eq.(\ref{eq:1.1}) without
(solid line) and with (dotted line) time reversal for a kick strength of
\(K=5.0\).}
\end{figure}

Let us now turn to our numerical comparison of the three cases outlined before. 
 In the quantum case the expectation value of the energy of the kicked rotator immediately after the n'th kick can be expressed by the formula
\begin{equation}
\langle E_n\rangle = K\sum^{n}_{m=1} \int^{2\pi}_{0} dq \rho_m (q) v_m (q) \sin q
\label{eq:1.5}
\end{equation}
This expression is valid for $g=0$ {\it and} $g\neq 0$. Here $\rho_m(q)$ and $v_m(q)$ are the probability distribution and the angular velocity of the BEC of kicked rotators in configuration space immediately after the m'th kick. They are defined by the macroscopic wavefunction 
$\psi_m(q)=\sqrt{\rho_m(q)} \exp (i \Phi_m (q))$, 
$v_m(q) = k^{\hspace{-2.4mm}-}\nabla\Phi_m(q)$. Expression (\ref{eq:1.5}) can be directly compared with the corresponding very similar formula for a classical ensemble of (noninteracting) kicked rotators
%
$\langle E_n\rangle = K \sum^{n}_{m=1}\overline{ p_m \sin q_m}$
%
where the horizontal bar denotes the average over the initial ensemble. 

In fig.\ref{fig2} the result for the ensemble of classical kicked rotators is shown. This figure is similar to a corresponding result in \cite{shep}. Along with the well-known diffusive increase of the mean energy over 500 kicking periods the result of a sudden time-reversal after 250 kicks is shown. Due to the chaoticity of the classical map in conjunction with the necessarily finite accuracy of the numerical simulation the dynamics is effectively irreversible.  
\begin{figure}
\includegraphics[width=7.0cm,height=4.5cm]{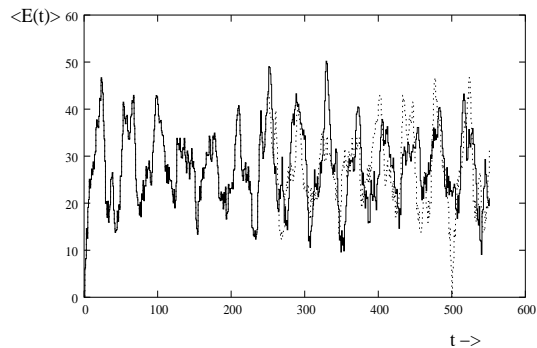}
\vspace*{0.1cm}
\caption{\label{fig3}Mean energy of the kicked rotator, according to the GP-equation (\ref{eq:1.4})
for \(g=0\), integrated with the split-operator method \cite{numerics}. Time reversal (dotted line)
\(\psi\to\psi^{*}\) is applied between the 250th and 251th kick}
\end{figure}
\begin{figure}
\includegraphics[width=7.0cm,height=4.5cm]{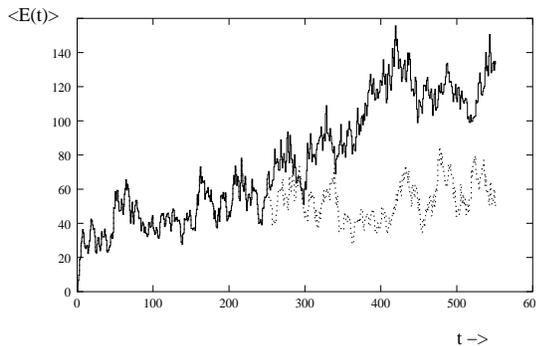}
\vspace*{0.1cm}
\caption{\label{fig4}Same calculation as for the kicked rotator of Fig. \ref{fig3},
but with interaction \(g=5.0\).}
\end{figure}
In fig.\ref{fig3} the corresponding result for the BEC of kicked rotators is plotted for the case where the interaction is negligible. Dynamical localization leads to the replacement of the classical diffusive increase of the mean energy by erratically and quasi-periodically oscillating interferences of the coherent matter wave, which are not averaged out due to the purity of the initial wavefunction. Furthermore, the time-reversal after 250 kicks leads to a complete retracing of the dynamics to the initial state \cite{shep}.

In fig.\ref{fig4} we finally plot the mean energy for a BEC of interacting kicked rotators. It is clear 
that the interaction has a strong quantitative but a much lesser qualitative influence on the interferences of fig.\ref{fig2}. It may be appropriate to stress at this point that figs.\ref{fig3} and \ref{fig4} were calculated with identical numerical codes, the only difference being the interaction constants $g=0$ for fig.\ref{fig3} and $g=5$ for fig.\ref{fig4}. (In both cases the angular momentum quantum number $l$ was restricted to the interval $-64 \le l \le 63$, and the $2\pi$-interval of $q$ was also discretized into 128 points appropriate for the fast Fourier transformation in a 64-Bit computation. 16,665 iterations of the split-operator method were carried out between two subsequent delta-kicks.) 
It is also apparent from fig.\ref{fig4} that the interacting dynamics becomes effectively irreversible again, even though the effective irreversibility can be seen to be much weaker than in the classical case. A similar remark applies to the diffusive increase of the mean energy. While it is clearly absent in fig.\ref{fig3} it reappears in fig.\ref{fig4}, however in a much weaker form and also in a more intermittent manner than in the classical case of fig.\ref{fig2}, which shows the strong influence of the coherence of the nonlinear matter wave. An increase in numerical accuracy (which must of course remain finite) only increases the time-interval over which the dynamics is reversible without changing the situation qualitatively.
The effective irreversibility and the diffusive increase of the mean
 energy both become stronger if the interaction strength $g$ is increased. This is because the $g|\psi |^2$-term introduces a rapidly varying $q$-dependent effective potential, which tends to wash out interference effects. 
An increase of $k^{\hspace{-2.4mm}-}$  
 works in the opposite direction, because the more pronounced quantization 
of angular momentum will enhance interference effects. 

In conclusion, we have considered a Bose-Einstein condensate of kicked rotators. We pointed out the existence of infinitely many explicitely time-dependent global conserved quantities of the underlying one-dimensional Gross-Pitaevskii equation on a ring constraining the dynamics, and gave a comparison of the mean energy of the Bose-Einstein condensate with and without interaction and the corresponding
classical ensemble. For fixed numerical accuracy we also compared the time-reversibilty of the simulation. To achieve a decent numerical accuracy,  a large number of time-steps {\it inbetween} the kicks is required in the case of nonvanishing interaction, where the dynamics cannot be reduced to a simple quantum-,map (see also \cite{raizen2}). As we showed here by the reversibility check this severely limits the number of kicks over which the dynamics is quantitatively reliable and hence reversible. It is tempting to interpret the lack of reversibility as another sign of chaos as in the classical case, but the existence of the Lax-pair and the conservation laws seem to call for caution. The experimental realization of the BEC of kicked rotators,
possibly along the line suggested here, both in the non-diluted interacting and in the diluted non-interacting case would be very interesting and might shed 
more light on the nature of the dynamics.

\begin{acknowledgments}
 This work was supported by the Deutsche Forschungsgemeinschaft within the
 SFB/TR 12. We enjoyed discussions with Konstantin Krutitsky.
 \end{acknowledgments}

%
%

%


\begin{thebibliography}{}
\bibitem{foot}  In ref.\cite{raizen2} a different realization of the kicks via impurities in a ring trap was proposed, which would, however, not as easily implement the $\delta$-kicks and the cosine-form of the potential.
\end{thebibliography}
\end{document}